\documentclass[pre,reprint, amsmath,amssymb,aps]{revtex4-1}
\usepackage{graphicx}
\usepackage{subfig}
\usepackage{amsmath}
\usepackage{amssymb}
\usepackage{enumerate}
\usepackage{color}
\usepackage[justification=raggedright,singlelinecheck=false,font=small]{caption}

\begin{document}

\title{Mechanisms to Splay-Bend Nematic Phases}

\author{N. Chaturvedi}
\email[]{nanditac@physics.upenn.edu}
\author{Randall D. Kamien}
\email[]{kamien@upenn.edu}
\affiliation{Department of Physics and Astronomy, University of Pennsylvania, Philadelphia,
PA, 19104-6396, USA}

\begin{abstract}

While twist-bend nematic phases have been extensively studied, the experimental observation of two dimensional, oscillating splay-bend phases is recent. We consider two theoretical models that have been used to explain the formation of twist-bend phases -- flexoelectricity and bond orientational order -- as mechanisms to induce splay-bend phases. Flexoelectricity is a viable mechanism, and splay and bend flexoelectric couplings can lead to splay-bend phases with different modulations. We show that while bond orientational order circumvents the need for higher order terms in the free energy, the important role of nematic symmetry and phase chirality rules it out as a basic mechanism. 

\end{abstract}

\maketitle
\section{Introduction}
\begin{figure}%
    \centering
    \subfloat[Twist-Bend]{{\includegraphics[width=2.0cm]{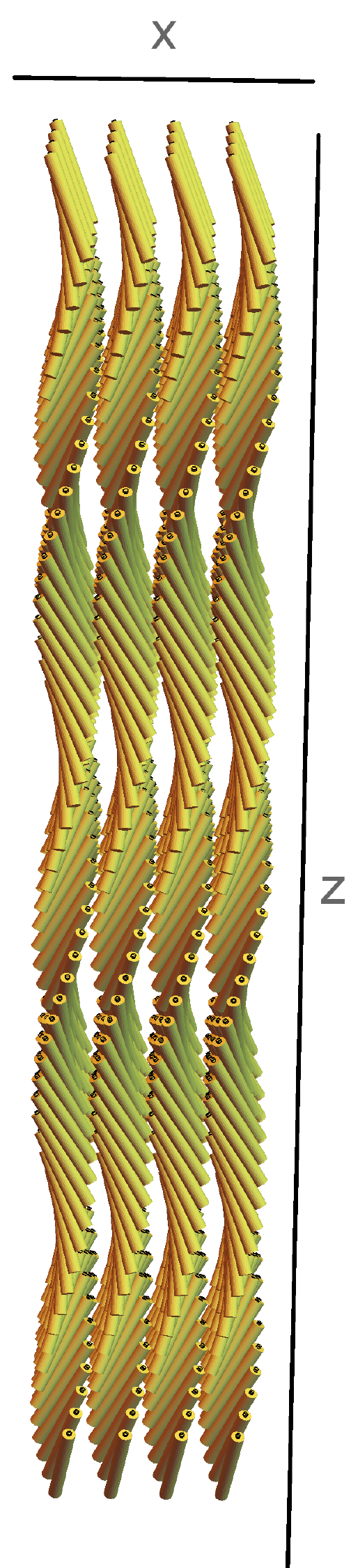} }}%
    \qquad
    \subfloat[Splay-Bend]{{\includegraphics[width=2.06cm]{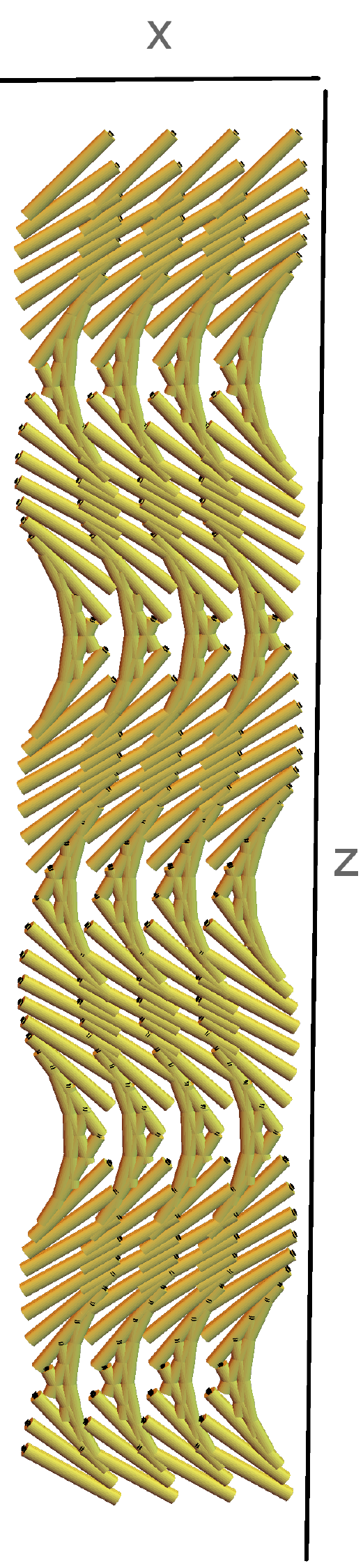} }}%
    \caption{The figure shows a twist-bend and splay-bend structure. the twist-bend structure has molecules rotating about the $z$ direction while maintaining a constant angle with it. In the splay-bend texture, molecules oscillate along the $z$ direction in two dimensions. \label{texture}}
\end{figure}

Liquid crystalline materials show a rich variety of structures and phases.  Indeed even if we focus on the smectic or cholesteric mesophases, there are a nearly unlimited variety of structures and motifs.  On the other hand, achiral nematic phases, the backbone of the display industry, the workhorse of experiment, and the most well understood have only a few variants (it has not escaped our attention that their simplicity is the key to their value as devices).  
Indeed, only a handful of distinct nematic phases have been found, and the space of possible configurations is highly restricted for achiral molecules. It is well known that achiral rod-like and discotic molecules form uniaxial nematics, and  also biaxial nematics \cite{madsen2004thermotropic, luckhurst2015biaxial, straley1974ordered}.   Over the past few decades, the study of bent core molecules has led to the discovery of a nematic phase in which the director field of achiral molecules follows an oblique helicoid, maintaining a constant oblique angle with a helical axis \cite{borshch2013nematic, henderson2011methylene, cestari2011phase, chen2014twist, mandle2014microscopy}. The texture is splay-free, having only twist and bend distortions. This new phase, the twist-bend phase, has attracted attention due to its unusual properties -- a spontaneously chiral phase is formed out of achiral molecules \cite{prasang2008spontaneous, meyer2015temperature}. Additionally, experiments show three times larger bend flexoelectric coefficients in bent core molecules than the typical value in rod-like liquid crystals \cite{harden2006giant, harden2010giant}. A schematic of this phase is shown in Fig. \ref{texture}.

With this phase as the backdrop, it is natural to contemplate additional nematic phases that show only twist and splay, or only splay and bend deformations. In this note we consider both bond orientational order and flexoelectricity as effects that can stabilize  ``splay-bend'' phases, also shown in the schematic in Fig. \ref{texture}. Although flexoelectricity has been considered before, we show that different forms of the splay and bend couplings can give us two distinct splay-bend phases with different modulations \cite{dozov2001spontaneous, mertelj2018splay}. The paper is organized as follows. In section \ref{bondsection}, we consider bond orientational order and find that nematic symmetry and phase chirality make bond order an unlikely mechanism for splay-bend. Next, in section \ref{instabilitysection} we consider flexoelectric effects and look at the splay and bend flexoelectric couplings that could give rise to splay-bend phases with different modulations.  In section \ref{splayandsbsection}, we look at the two different `splay' phases that have been addressed in the literature, splay-bend \cite{dozov2001spontaneous} and splay nematic phases \cite{mertelj2018splay}, and show that these are related to each other by an exchange of the bend and splay deformations.

The mechanism behind the emergence of the twist-bend and splay-bend phases remains debated. Initial work argued that a purely elastic instability,  resulting from negative bend elastic constants, could explain the emergence of both these phases \cite{dozov2001spontaneous, memmer2002liquid, meyer2016local}. However, this leads to a free energy unbounded from below -- higher order and degree terms are necessary to find stable extrema. 
More recent theoretical work shows that a linear coupling between polar order and the deformations of the nematic director can give effective elastic constants, which can then be driven negative with changing temperature \cite{shamid2013statistical, meyer2013flexoelectrically}. Bend flexoelectric couplings have been used to explain twist-bend phases, and a combination of both bend and splay flexoelectricity to explain splay-bend phases. Recent work shows that combinations of flexoelectricity and intrinsic chirality also predict yet unseen, but related, modulated phases \cite{longa2016modulated}.

Another mechanism that does  not require higher order terms does exist for the twist-bend texture \cite{kamien1996liquids} but requires chiral bond order: upon cooling, nematic liquid crystals can give rise to a liquid crystalline phase with nematic order and hexatic order in the plane perpendicular to it \cite{toner1983bond}.  If the hexatic order is itself chiral, then the twist-bend texture is stable.  Such a mechanism would predict the emergence of twist-bend and splay-bend phases without the need for stabilizing arbitrary higher order terms but pushes the problem on to find a mechanism for spontaneous achiral symmetry breaking in the case of achiral molecules.

\section{Bond Orientational Order}\label{bondsection}

Previous work shows how hexatic bond order in a chiral liquid crystal can give rise to a twist-bend phase, while circumventing the need for higher order terms  \cite{kamien1996liquids}. We consider now whether this is a viable mechanism to induce the splay-bend phase. Consider a nematic system with bond orientational order in the plane perpendicular to the nematic director. For our purposes, it is sufficient to consider the general case without specifying the number of nearest neighbors. 

The fluctuations in the nematic director, $\bf{n}$, are given by the Frank free energy density,

\begin{eqnarray}
f_{\mathbf{n}}&=&\frac{K_1}{2} \left[\mathbf{n}(\nabla \cdot \mathbf{n} )\right]^2 +\frac{K_2}{2}\left[ \mathbf{n}\cdot (\nabla\times \mathbf{n})\right]^2 \nonumber\\&&\qquad + \frac{K_3}{2}\left[(\mathbf{n}\cdot\nabla)  \mathbf{n}\right]^2 \label{frankenergy}
\end{eqnarray}

\noindent where $K_1, K_2$ and $K_3$ are the splay, twist and bend elastic constants, respectively. Here and throughout we require that these elastic constants are positive.  Apart from the contributions to the free energy from modulations in the director field, we want to account for interactions between the director and the bond angle. The bond angle, $\Phi$, quantifies the bond order in the system. The definition of $\Phi$ depends on the definition of the nematic director field \cite{kamien1996liquids}. In particular, it follows the nematic symmetry, and $\Phi \rightarrow -\Phi$ under the transformation $\mathbf{n}\rightarrow -\mathbf{n}$. We expect that the bond order contribution to the free energy density has a term that penalizes any sharp changes in $\Phi$, and a term that captures the interaction between $\Phi$ and $\mathbf{n}$. 

Since we require that the overall nematic symmetry is preserved in the free energy density, any term that represents the interaction between the bond angle and the nematic director must have an even power of $\Phi$ and $\mathbf{n}$ together. This means, for a term linear in $\nabla \Phi$, the interaction term must have an odd power in $\mathbf{n}$.

The twist-bend phase has a chiral structure, and so a chiral interaction term is expected. In order to construct the interaction term then, we want a vector with an odd number of derivatives to account for chirality, and an odd power of $\mathbf{n}$ to preserve nematic symmetry. The lowest order term that satisfies these constraints is $ \mathbf{n} \cdot \nabla \Phi $.   Considering this term,

\begin{equation}
f_{\Phi}=\frac{K_A}{2}(\nabla\Phi)^2-K_A q_0( \mathbf{n}\cdot\nabla\Phi)
\end{equation}

\noindent where the full free energy density is $f=f_{\mathbf{n}} + f_\Phi$ and $f_\Phi$ is the contribution to the free energy density from bond orientational order. The total free energy can be minimized to determine the parameters of the phase and the bond angle, $\Phi$, as a function of the Frank constants, the pitch ($q_0$), and the bond-angle stiffness ($K_A$). Since the interaction term is of lower order than the terms in the Frank elastic energy, the total free energy remains bounded from below.   Indeed, extremizing over $\Phi$ we have $\nabla^2\Phi=q_0\nabla\cdot\mathbf{n}$.  For the twist-bend texture $\mathbf{n}_{\rm{tb}}=[\cos(qz)\cos\theta,\sin(qz)\cos\theta,\sin\theta]$ and we can only have $\nabla\Phi=v_0\hat z$, a constant vector along the $z$-axis.  Minimizing over the value of $\mathbf{v}_0$ and integrating over a period generates a term \cite{kamien1996liquids}
\begin{equation}
f_{\Phi}= -\frac{K_Aq_0^2}{2}\sin^2\theta
\end{equation}
and the bond order acts as a magnetic aligning field as studied half a century ago by R.B. Meyer \cite{meyer1969distortion}, stabilizing the texture.

\begin{figure}%
    \centering
      \subfloat[Splay-bend phase in Eq. (\ref{ansatz1})]{{\includegraphics[width=3.5cm]{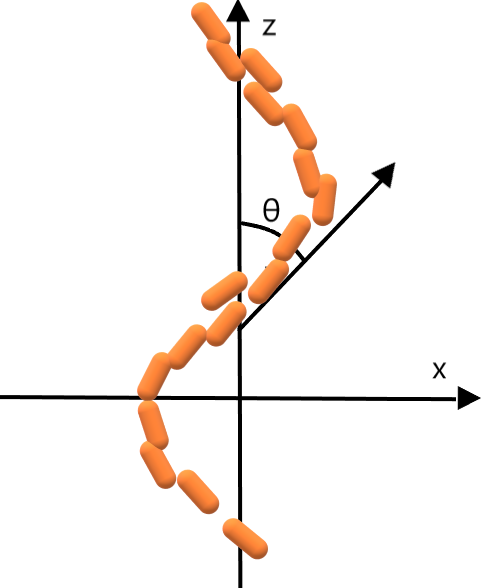} }}%
        \qquad
      \subfloat[Splay-bend phase in Eq. (\ref{ansatz2})]{{\includegraphics[width=3.5cm]{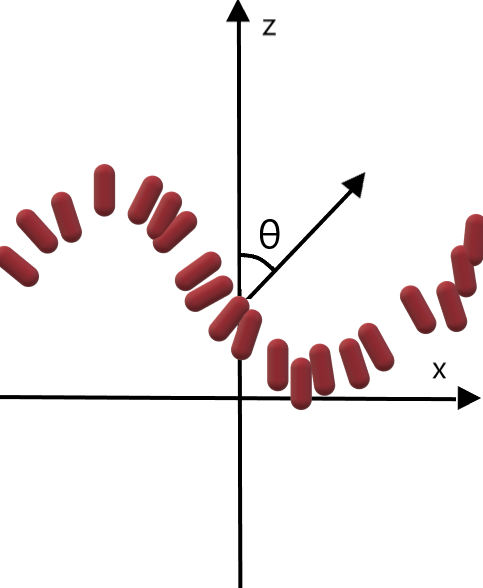} }}%
    \caption{Schematics of the splay-bend phase in the \textit{ansatz} in Eqs. (\ref{ansatz1}) and (\ref{ansatz2}) showing the maximum angle $\theta$. The modulations are parallel to the average nematic director direction  in the left schematic, and perpendicular to it in the right.\label{sbschematic}}
\end{figure}

Following a similar argument, we might consider the possibility that bond orientational order is also a mechanism for the formation of splay-bend phases. For splay-bend, we use the \textit{ansatz},
\begin{equation}
\mathbf{n}_1=\Big\{\sin\left[\theta \sin (q z)\right],0,\cos\left[\theta \sin (q z)\right]\Big\} \label{ansatz1}
\end{equation}
This describes an oscillating, two-dimensional structure that alternates between regions of splay and bend deformations. Here, $q$ is the pitch of the phase and $\theta$ is the maximum angle to which the molecules tilt. The direction of modulation here is parallel to the average nematic director field as shown in Fig. \ref{sbschematic}.

%

Using this to calculate the splay and bend free energy density contributions, and averaging over a period $\pi/q$, we find a free energy density

\begin{align}
\bar f_{\mathbf{n}_1}=\frac{K_1 q^2 }{8}\Big[\theta^2-\theta J_1(2\theta)\Big]
+\frac{K_3 q^2}{8}\Big[\theta^2+\theta J_1(2\theta)\Big]
\label{splaybendz}
\end{align}

 \noindent Here, $J_\nu(z)$ are Bessel functions of the first kind.  Using the properties of Bessel functions it is straightforward to check that when both $K_1$ and $K_3$ are positive, $\bar f_{\mathbf{n}_1}\ge 0$ and has only one minimum at $\theta=0$, the uniaxial namatic.   Since the  splay-bend phase is achiral an achiral coupling is necessary, $\tilde f_\Phi$. 
The symmetries that the new term must have are as follows: continuing to require that the nematic symmetry, $\mathbf{n}\rightarrow -\mathbf{n}$, is preserved, the interaction term must have an even power of $\Phi$ and $\mathbf{n}$ together. Further, since the texture is achiral, we assume that the interaction term must also be achiral and thus even in derivatives of fields. Thus a term linear in $\nabla\Phi$, requires a vector with an odd order of derivatives, and an odd power of $\mathbf{n}$.

We may then list our the possibilities for the lowest order term: one could consider interactions that involve the splay vector, $\mathbf{n}(\nabla\cdot \mathbf{n})$, but these do not follow the nematic symmetry. The same is true for interactions that involve the bend vector, $\mathbf{n}\times (\nabla\times \mathbf{n})$. 

One possibility that has the required symmetries is $\nabla \Phi\cdot (\nabla\times \mathbf{n})$. In this case, the extremal equation for $\Phi$ is again $\nabla^2\Phi=0$.  Since $\nabla\times\mathbf{n}_1 = [q\theta\cos(qz)\sin\left(\theta\sin(qz)\right),0,0]$, we can consider the standard harmonic solutions of Laplace's equation for $\Phi$.  If $\nabla\Phi\cdot\nabla\times\mathbf{n}_1\ne 0$ then $\Phi$ must depend on $x$.  There is the solution linear in $x$ which, when inserted and averaged over a $z$ period results in no coupling between the bond order and the director.  Other solutions are of the form $\cosh(\alpha x_i)\cos(\alpha x_j)$ where $i\ne j$ and $\alpha$ is a constant.  Since we would need $\partial_x\Phi\ne 0$, the only possible term that would not vanish upon spatial averaging would be of the form $\Phi=\cosh(\alpha x)\cos(\alpha z)$ (up to translations).  Unfortunately, while surviving the $z$-averaging, a solution like this would lead to an unbounded free energy density.  Whether it is possible to have defect walls between regions of bounded $\nabla\Phi$ is the topic of future work.

Finally, were we to consider an interaction higher order in derivatives than either the  bend or splay vectors, we would generate an odd power of $q$ higher than $2$ in the free energy integrated over one pitch, requiring even higher order terms to assure stability.  Since that was the {\sl raison d'\^{e}tre} for considering this mechanism, we conclude that there are then no interaction terms that have the appropriate symmetries, and a low enough order to give a non-trivial minimum for $q$ and $\theta$. 

We conclude then, that bond orientational order is not a simple mechanism that can give splay-bend phases. In order to get a splay-bend phase, a vector field, like the polarization vector, $\mathbf{P},$ is required \cite{shamid2013statistical}. Such a field plays the part of a vector that need not follow the nematic symmetry. Several of the interaction terms that are not available to us with the bond angle are then permitted by symmetry.

\section{Flexoelectricity}\label{instabilitysection}

Recall that the flexoelectric effect is a linear coupling between a polarization vector and director deformations. A coupling may be constructed with either the splay or bend vectors that, in turn, gives rise to an effective negative $K_1$ or $K_3$, respectively \cite{shamid2013statistical}. Such a coupling induces spontaneous splay or bend in the system. Previous work has shown how a negative effective $K_3$ can lead to both the twist-bend and splay-bend phases \cite{dozov2001spontaneous}. Similarly, a negative effective value of $K_1$ has been used to explain the observation of the splay nematic phase \cite{mertelj2018splay}.
 
We look at both of these couplings independently. We consider the following {\sl ansatz} for the polarization vector, $\mathbf{P}$ and nematic director field $\mathbf{n}$ \cite{mertelj2018splay},

\begin{eqnarray}
\mathbf{n}_2&=&\Big\{\sin\left[\theta \sin (q x)\right],0,\cos\left[\theta \sin (q x)\right]\Big\} \\
\mathbf{P}&=& \mathbf{n}_2 p \cos q x \label{ansatz2}
\end{eqnarray}

\noindent This \textit{ansatz} is different from the one in Eq. (\ref{ansatz1}), and the direction of modulation is perpendicular to the average direction of the nematic director field, as is shown in Fig. \ref{sbschematic}. We will show in the next section how the two splay-bend systems can be mapped on to each other.  This form for the polarization, $\mathbf{P}$, breaks the nematic up/down symmetry.   When averaging over the sample, modulations that are at different wavelengths than $2\pi/q$ will vanish and so we pick the dipole modulation accordingly.
For a splay flexoelectric coupling, the free energy density is

\begin{equation}
f_{\hbox{\tiny splay}}= f_{\mathbf{n}}-\gamma \mathbf{P}\cdot \left[ \mathbf{n} (\nabla\cdot \mathbf{n})\right]+\frac{b}{2} (\nabla \mathbf{P})^2+\frac{t}{2}  \mathbf{P}^2
\label{equation}
\end{equation}

\noindent where $\gamma$, $b$ and $t$ are Landau coefficients. The value of $t$ changes with temperature and drives the transition to a spontaneously polarized state \cite{mertelj2018splay}. Since the free energy is second order in $\mathbf{n}$, the effective period of its variation is $\pi/q$. Inserting the \textit{ansatz} into the free energy density, and integrating over a period $\pi/q$, we find an average free energy density of

\begin{align}
\bar f_{\hbox{\tiny splay}}&= \frac{K_1}{8} q^2 \left[ \theta ^2 +  \theta  J_1(2 \theta )\right]+\frac{K_3}{8} q^2 \left[  \theta ^2 -  \theta   J_1(2 \theta )\right]\nonumber\\&\qquad- \gamma  p q \frac{J_1(\theta )}{\theta}+\frac{t  p^2 }{4 }+ \frac{b}{16}  p^2 q^2  \left(3 \theta ^2+4\right)
\end{align}

\noindent This free energy can then be minimized with respect to $p$ and $q$ to obtain the following expressions at the free energy minimum,

\begin{eqnarray}
q_{splay}^2&=&-\frac{4 \sqrt{t}}{b(3  \theta ^2+4 )} \times  \label{qsquared} \\&&\left\{\sqrt{t}+\frac{ 2 \sqrt{2} \gamma | J_1(|\theta| )|  }{\sqrt{ (K_1+K_3)\theta^2 +  \theta(K_1-K_3) J_1(2 \theta )} }\right\} \nonumber\\
p_{splay}&=&\frac{8 \theta \gamma  J_1(|\theta|)  q_{splay}}{ |\theta| \left( b\left(3 \theta ^2+4\right) q_{splay}^2+4 t\right)}
\end{eqnarray}

\noindent  Note that the radicand in (\ref{qsquared}) is non-negative ($\bar f_{\mathbf{n}_1}\ge 0$).   In order to check the validity of these expressions, we plug in typical values of the different constants and take $\theta\sim 1$. Using $ K_1= 10  \, \text{pN},  K_3=1  \,\text{pN}, \gamma=10^{-3}  \, \text{ V}, b=2 \times 10^{-18}  \, \text{V m}^3/(\text{A s})$ and $t=8 \times 10^{-8}  \,\text{V m/(A s)}$, we obtain $q= 0.1  \,\text{nm}^{-1}$ and $p= 10^7 \, (\text{A s})/\text{m}^2$. This is consistent with experiments where a nanometer range for pitch is observed  \cite{mertelj2018splay}. Further, using the typical density of $1 \,\text{g}/\text{cm}^3$, the value of the polarization density, $p$ translates to a molecular polarization of $10 \, \text{Debye}$, which is approximately the same as that of the molecule of RM734 seen to form splay-bend phases \cite{mertelj2018splay}.

\begin{figure}[]
\centering
\includegraphics[width=0.5\textwidth]{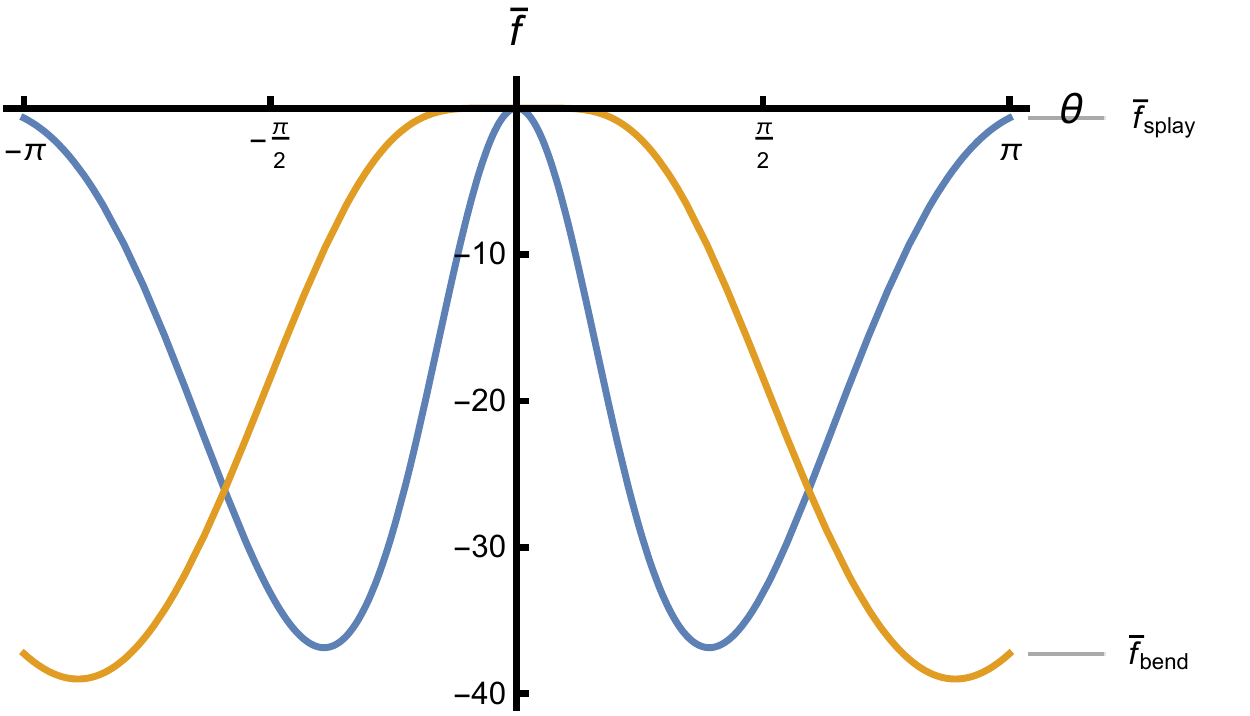}
\caption{Plot of $\bar f_{splay}$ and $\bar f_{bend}$ as a function of $\theta$ at the free energy minimizing values of $q$ and $p$. The plots clearly show that both free energies have a minimum at a non trivial value of $\theta$, implying that the splay-bend phase is a possibility with both couplings. The parameter values are $ K_1=1.5,  K_3=2, \gamma=40, b=2$ and $t=10$. \label{freeenergyplot}}
\end{figure}

We substitute these expressions for $p_{splay}$ and $q_{splay}$ into the free energy and plot it as a function of the maximum angle $\theta$ in Fig. \ref{freeenergyplot}. As can be seen, there is a nontrivial minimum at a non-zero value of $\theta$, so the splay-bend phase is stable in the case of a splay flexoelectric coupling.

In the case of a bend flexoelectric coupling, the only term that changes is the interaction term with coupling $\gamma$.  A bend flexoelectric coupling is of the form  $ \mathbf{P}\times \left[ \mathbf{n}\times (\nabla\times \mathbf{n})\right]$.  However, this is a vector.  If, however, the material were sandwiched between two different plates separated in the direction perpendicular to director (the $y$-axis), then a coupling of the form $\hat y\cdot\left( \mathbf{P}\times \left[ \mathbf{n}\times (\nabla\times \mathbf{n})\right]\right)$ is allowed.  In this case the average free energy density is
\begin{eqnarray}
\bar f_{\hbox{\tiny bend}}&=& \frac{K_1}{8} q^2 \left[ \theta ^2 +  \theta  J_1(2 \theta )\right]+\frac{K_3}{8} q^2 \left[ \theta ^2 -  \theta   J_1(2 \theta )\right]\nonumber \\ &&\quad- \gamma  p q \pmb{H}_1(\theta )+\frac{t  p^2 }{4 }+ \frac{b}{16} p^2 q^2  \left(3 \theta ^2+4\right)
\end{eqnarray}

\noindent Here, $\pmb{H}_\nu(z )$ is the Struve function of order $\nu$. Repeating the same procedure as earlier, we plot the average free energy density in Fig. \ref{freeenergyplot}. As can be seen, a nontrivial minimum exists at a higher value of $\theta$ than for splay flexoelectricity. Thus, we conclude that the splay-bend phase given by the \textit{ansatz} in Eq. (\ref{ansatz2}) can be obtained by either splay flexoelectric coupling or a bend flexoelectric coupling along with a sample asymmetry, providing the direction $\hat y$.

\section{Splay-Bend and Splay Nematic Phases}\label{splayandsbsection}

Previous work on nematic phases with splay and bend modulations makes a distinction between the \textit{ansatz} in Eq. (\ref{ansatz2}), a `splay nematic phase', and the `splay-bend phase' in Eq. (\ref{ansatz1})  \cite{mertelj2018splay}. In particular, the direction of the modulation is perpendicular to the director in Eq. (\ref{ansatz2}), as opposed to along the director, as in Eq. (\ref{ansatz1}). In the `splay-bend phase', the splay and bend contributions to the free energy density, integrated over a period $\pi/q$, are then,

\begin{equation}
\bar f_{\mathbf{n}_2}=\frac{K_1 q^2 }{8}\Big[\theta^2+\theta J_1(2\theta)\Big] +\frac{K_3 q^2 }{8}\Big[\theta^2-\theta J_1(2\theta)\Big]
\end{equation}

\noindent As can be seen from a comparison of the above equation with Eq. (\ref{splaybendz}), the splay and bend contributions have been interchanged. The two systems can be mapped on to each other by exchanging $K_1$ with $K_3$. The `splay nematic phase' and the `splay-bend phase' are closely related phases. This is expected since a rotation of the nematic director field by $\pi/2$, as would be required to turn $\mathbf{n}_1$ into $\mathbf{n}_2$, would turn splay deformations into bend deformations and bend into splay.

We could have begun by using the \textit{ansatz} in Eq. (\ref{ansatz1}), and repeated the process outlined in section \ref{instabilitysection} by inserting the new \textit{ansatz} into the free energies with the two different flexoelectric couplings. Minimizing with respect to $q$ and $p$, we would find that the results in Section \ref{instabilitysection} are reversed, and the curves for $\bar f_{\hbox{\tiny splay}}$ and $\bar f_{\hbox{\tiny bend}}$ interchanged in Fig. \ref{freeenergyplot}. Thus, both the splay-bend and `splay nematic' phases can be obtained with splay and bend flexoelectric couplings, and can be related to each other by an exhange of the bend and splay elastic constants.

\section{Conclusions}
We have demonstrated that while bond orientational order is a mechanism that could circumpass the problem of an unbounded free energy, nematic symmetry and achirality of the splay-bend phase prevent it from explaining the formation of the splay-bend phase. Flexoelectricity provides a viable mechanism for introducing splay and bend modulations in nematic systems. Bend and splay flexoelectric couplings lead to effective elastic constants that stabilize splay and bend modulations. Both flexoelectric couplings can give rise to splay-bend phases with modulation in the average direction of the director field or modulations perpendicular to the average direction of the nematic director. These two modulations, treated previously in the literature as `splay-bend' and `splay-nematic' phases, are related to each other by  an exchange of the splay and bend elastic constants.

\acknowledgments
We thank A. Mertelj for helpful discussions.  This work was supported by a Simons Investigator grant from the Simons Foundation to R.D.K. and NSF DMR12-62047.


%
%

%


\bibliography{bendsplay}

\end{document}